# A Robust Repetitive Controller for Fast AFM Imaging

Serkan Necipoglu, Selman A. Cebeci, Y. Emre Has, Levent Guvenc, and Cagatay Basdogan

***Abstract*—Currently, Atomic Force Microscopy (AFM) is the most preferred Scanning Probe Microscopy (SPM) method due to its numerous advantages. However, increasing the scanning speed and reducing the interaction forces between the probe's tip and the sample surface are still the two main challenges in AFM. To meet these challenges, we take advantage of the fact that the lateral movements performed during an AFM scan is a repetitive motion and propose a Repetitive Controller (RC) for the *z*-axis movements of the piezo-scanner. The RC utilizes the profile of the previous scan line while scanning the current line to achieve a better scan performance. The results of the scanning experiments performed with our AFM set-up show that the proposed RC significantly outperforms a conventional PI controller that is typically used for the same task. The scan error and the average tapping forces are reduced by 66% and 58%, respectively when the scan speed is increased by 7-fold.**

## I. Introduction

THE use of AFM in many different applications has increased rapidly since its invention in 1986 [1]. It has several advantages over the other scanning microscopy methods such as SEM (scanning electron microscopy) and STM (scanning tunneling microscopy). These include easy sample preparation, the ability to scan surfaces in air, liquid, and vacuum environments and relatively lower cost. For this reason, AFM has a wide range of use in material science, electronics, optics, semiconductor industry, biology and other areas of life sciences. The primary components of an AFM setup operating in dynamic mode are illustrated in Fig. 1. The sample to be scanned is placed on a three degrees-of-freedom piezo-actuated stage. In tapping mode AFM operation, first, the scanning probe is excited to vibrate sinusoidally near the resonance frequency in free air and then brought close to the sample to lightly tap its surface. In amplitude modulation scheme, the tapping amplitude of the probe, $A_{act}$, is kept at a set value, $A_{set}$, by a feedback controller adjusting the vertical (*z* axis) movements of the stage while the scan proceeds on the lateral axes (*x* and *y*). This lateral motion is either triangular or quadratic and controlled separately. Typically, PID type controllers have been used for controlling the vertical and lateral movements of the stage [2]. The movement of the stage along the vertical axis (*z*-axis) is recorded as the surface height for each grid point on the *x*-*y* plane to construct a topographic map of the scanned surface.

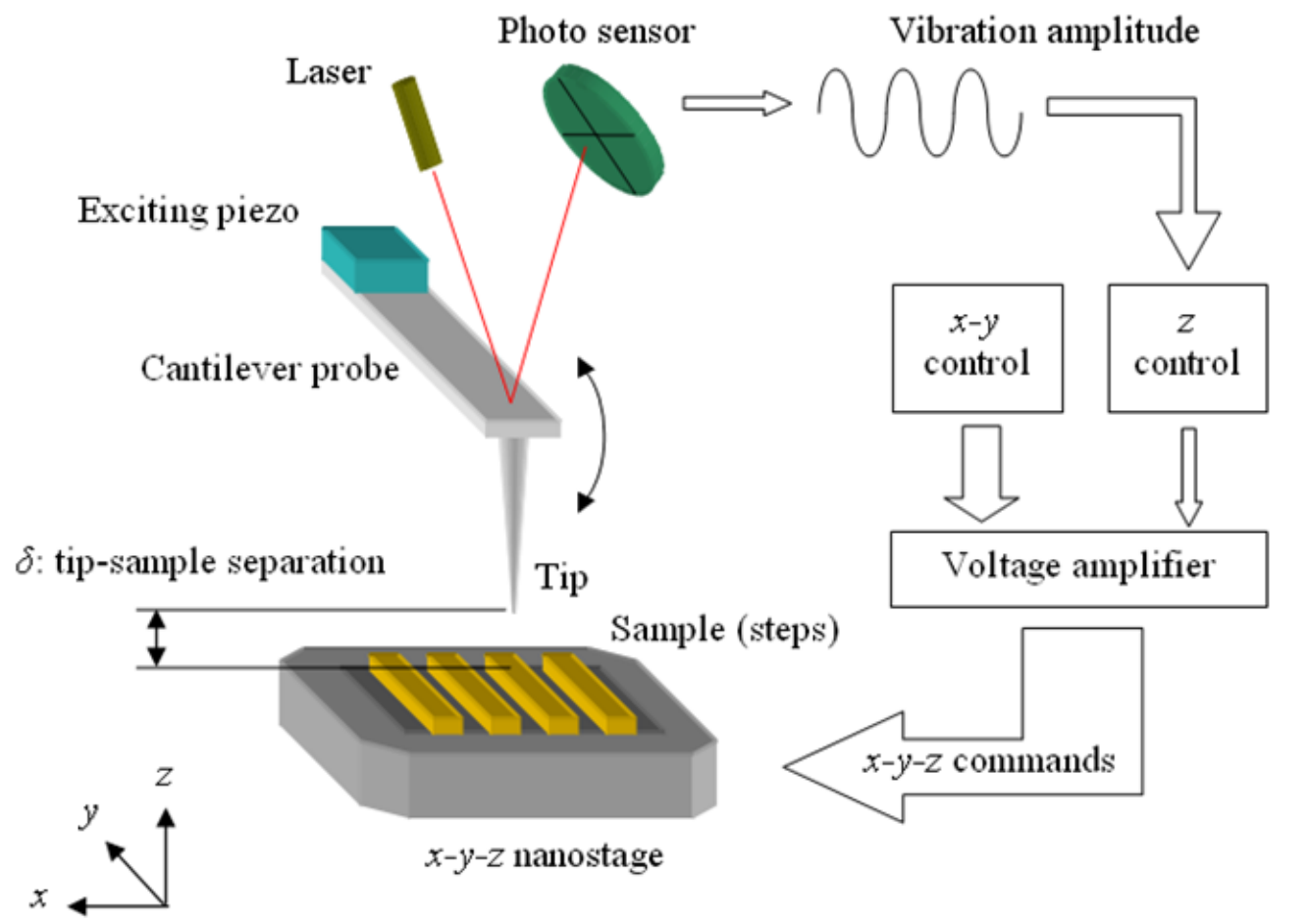

Fig. 1. Dynamic-mode AFM: The probe is excited by a piezoelectric element to vibrate sinusoidally near the resonance frequency and its vibration amplitude is measured by a laser beam reflected on a photo sensor. The sample to be scanned is placed on a piezo-actuated nano-stage. Typically, the controller used for the lateral scan motion on *x-y* plane is open loop, while a closed loop controller is used for the *z* motion to acquire the surface topography based on the vibration amplitude of the probe.

Increasing the scanning speed and reducing the magnitude of the interaction forces between the probe tip and the sample surface are two main challenges in AFM scanning. For example, it takes several minutes to scan even a small area of 2 μm by 2 μm. Besides, AFM is too slow to capture the phases of some rapid biological phenomena and new high-speed imaging techniques (see the review in [3]) have already shown

S. Necipoglu is with MEKAR (Mechatronics Research Labs) and Automotive Control and Mechatronics Research Center, Department of Mechanical Engineering, Istanbul Technical University, Istanbul, 34437, Turkey, e-mail: necipoglu@itu.edu.tr.

S. A. Cebeci is with Robotics and Mechatronics Laboratory, College of Engineering, Koc University, Istanbul, 34450, Turkey, e-mail: secebeci@ku.edu.tr.

Y. E. Has is with Robotics and Mechatronics Laboratory, College of Engineering, Koc University, Istanbul, 34450, Turkey, e-mail: yhas@ku.edu.tr.

L. Guvenc is with MEKAR (Mechatronics Research Labs) and Automotive Control and Mechatronics Research Center, Department of Mechanical Engineering, Istanbul Technical University, Istanbul, 34437, Turkey, e-mail: guvencl @itu.edu.tr.

C.Basdogan is with Robotics and Mechatronics Laboratory, College of Engineering, Koc University, Istanbul, 34450, Turkey, phone: (90+) 212 338 1721, fax: (90+) 212 338 1548, e-mail: cbasdogan@ku.edu.tr.

to be highly effective [4], [5]. Moreover, during the scan process, the sample and the probe can be damaged easily if the magnitude of the interaction forces is high. This is especially critical when scanning biological samples since high interaction forces can cause an irreversible damage on the sample**.**

One of the major limitations on the scan speed of an AFM is the bandwidth of the piezo-actuated stage. The physical bandwidth is limited by the resonance frequency of the mechanical structure, where the control bandwidth is limited by the gain margin of the closed-loop system commanding the stage. One way to obtain a higher scanning speed is to improve the mechanical design of the stage [6]. The other is to extend the control bandwidth by using a more sophisticated controller instead of the conventional PID controller. For example, feedback controllers based on $H_\infty$ theory are proposed in [7] and [8] for adjusting the movements of the stage in the *z*-axis and the lateral axes. A scanning speed that is five times higher than that of the conventional PID controller is reached in [7]. An inversion-based feed-forward controller is presented in [9] for improving the tracking performance of the stage in lateral axes at high frequency. A detailed comparison of various feedback and/or feed-forward controllers used for the lateral motion is given in [10]. In addition to adapting a more sophisticated controller, one can also take advantage of the fact that the successive scan lines are quite similar to each other to further improve the controller performance. For example, a PID feedback controller combined with a feed-forward is presented in [11], where the profile of the previous scan line is supplied to the feed-forward controller to improve the scan speed. In [12], a more sophisticated $H_\infty$ feedback and feed-forward controllers are used in tandem for the same purpose. In [13], a surface topography learning observer is designed by using a rule-based approach for perfect tracking control (PTC) in vertical motion. The tracking error is improved by up to 80%. An iterative learning controller (ILC) proposed in [14] improves the tracking performance of a contact mode AFM in vertical axis by 8-fold at high scan speeds. In [15], an RC is designed and implemented for controlling the lateral scan motion in AFM.

In this paper, we propose an RC for dynamic-mode AFM, which can reject the repetitive disturbances (i.e. surface profile) successfully. This is due to the fact that the memory loop inside the RC becomes the generator of any repetitive input signal when the period of the memory loop is adjusted to match the period of the repetitive input signal [16]. Based on the internal model principle, the generator of the input signal on the forward path of a feedback loop, drives the steady state error to zero, and provides perfect tracking [17]. From another point of view, the RC introduces infinite loop gains at the fundamental frequency of the input signal and its harmonics [18], which is desirable, but may cause instabilities, for example, in some of the methods mentioned above. However, RC provides robust stability when it is used with appropriate filters [19] – [21].

The proposed RC is implemented in tapping mode AFM and tested by our home-made AFM set-up. Its performance is compared to that of the conventional PI controller. It is proposed as an add-on controller to a conventional PI controller. Hence, it can be turned on and off whenever necessary during the scan since its operation does not depend on the initial conditions as it does in ILC [22]. The main contributions of this study are itemized below:

1) In our implementation, we control the *z*-axis movements of the stage using an RC. The proposed RC increases the scan speed by improving the control bandwidth of the system while maintaining the stability and the image quality.
2) In addition to improving the tracking performance, we show that the RC also reduces the interaction forces between the probe tip and the sample surface. Lower tapping forces help preserve the sample and lead to the longer use of the scanning probe.
3) We propose a new iterative approach for setting the tunable parameters of an RC in experimental settings. Using this approach, one can successfully determine the optimal parameters of an RC, resulting in better performance in AFM scanning than that of a conventional PI controller.

The following section provides the necessary background on the discrete-time implementation of an RC for tapping mode AFM. The components of our AFM set-up are introduced in Section III. In Section IV, we identify the dynamical characteristics of our stage controlled by the RC. The design and the analysis of the RC filters for the stage are given in Section V. The scanning experiments and the results are reported in Section VI. Finally, we conclude the study in Section VII with a discussion of the performed work.

## II. Repetitive Control

In Fig. 2a, the RC assisted PI control scheme for a tapping mode AFM is presented in the discrete time domain. The probe which appears in the diagram acts like a sensor, measuring the separation distance, $\delta$, between the probe tip and the sample surface. Its dynamic response is approximately linear and much faster than that of the stage. Hence, it is considered to be a static device with a gain value equal to unity (i.e. $P(z) = 1$) in our design. The nonlinearities in the probe dynamics and the ones due to the interactions between the probe tip and the sample surface are treated as unstructured uncertainties in our approach, which are handled by the robust design of RC. The surface topography and the initial distance between the tip and the sample are treated as disturbances, $d$, to be rejected by the controller. The $q$ filter in the memory loop, which is a low-pass filter with a DC gain equal to 1, is required to filter out the infinite loop gains that are introduced by the memory loop at the high frequency harmonics. Basically, it defines the control bandwidth of the RC and also prevents the excitation of the undesired dynamics at high frequencies. The $b$ filter helps to maintain the inequality condition given in (1), which is sufficient for the stability of the RC system. The function $R(\omega)$ is called the regeneration spectrum [23].

$$R(\omega) = \left| q(z) \left[ 1 - b(z) \frac{PI(z)G(z)}{1+PI(z)G(z)} \right] \right| < 1 \quad where \quad z = e^{j\omega T}. \tag{1}$$

Here, the frequency of interest varies from $0 < \omega < \omega_{max}$, where $\omega_{max} = \pi/T$. The sampling time $T$ is typically selected as 5-10 times smaller than the period of the highest frequency of interest. The stability condition given in (1) encourages the selection of the $b$ filter as simply the inverse of $PI(z)G(z)/[1+PI(z)G(z)]$ in order to make $R(\omega)$ as small as possible.

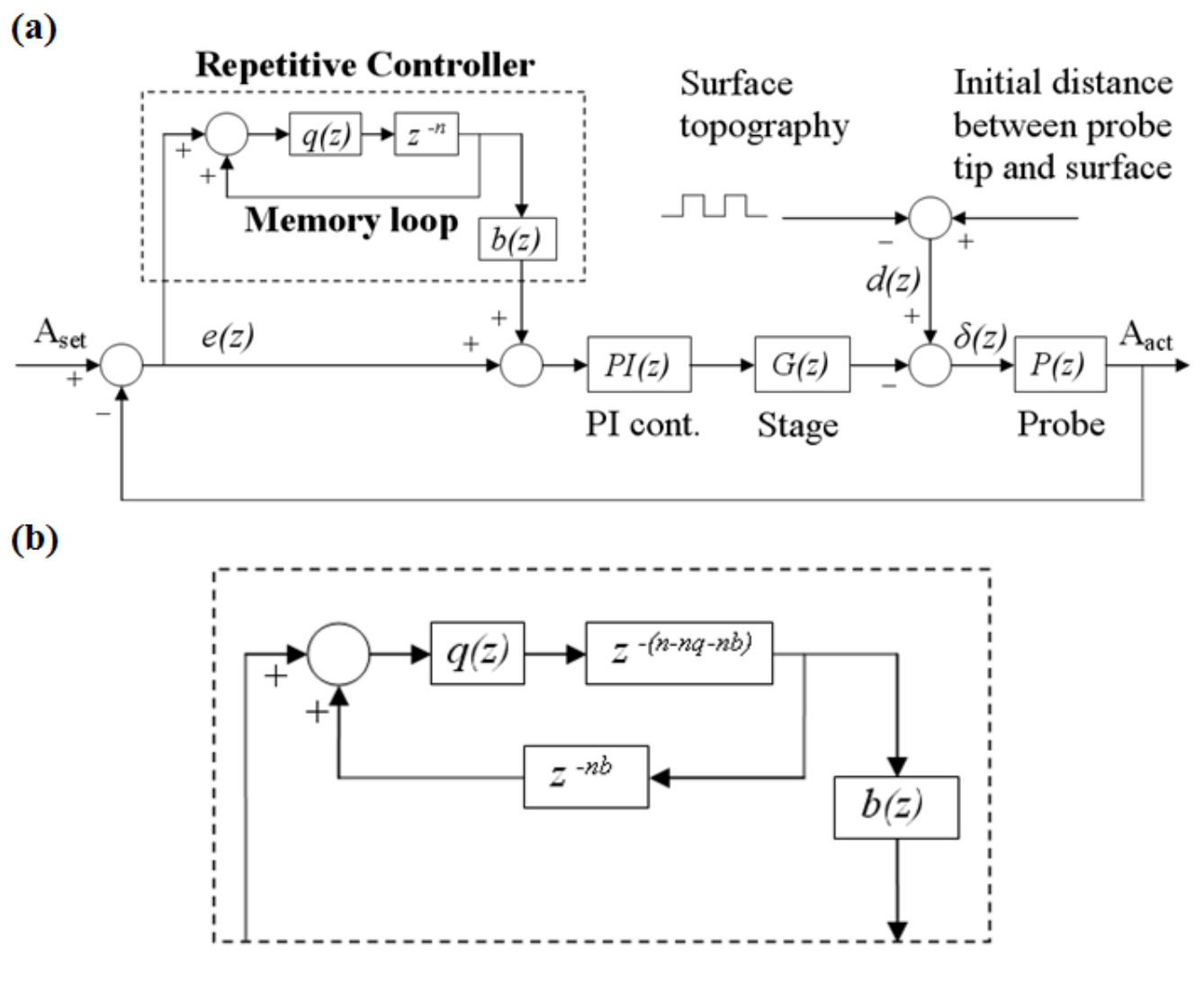

Fig. 2. (a) The RC scheme for a tapping mode AFM. (b) The modified RC structure for incorporating the sample advances.

In order to improve the performance of the RC, a small sample advance, $z^{nq}$, is incorporated into the $q$ filter to cancel out its negative phase. Similarly, a small sample advance, $z^{nb}$, is incorporated into the $b$ filter to cancel out the negative phase of $PI(z)G(z)/[1+PI(z)G(z)]$ and consequently help to maintain $R(\omega)<1$. Hence, the filters with sample advances can be written as

$$q'(z) = z^{nq} q(z) \quad and \quad b'(z) = z^{nb} b(z). \tag{2}$$

These small sample advances can easily be absorbed in the much larger period of the exogenous input signal and do not cause a problem in implementation if the memory loop is modified as shown in Fig. 2b. Although the values of $nb$ and $nq$ can be estimated from the phase diagrams of the filters directly, further adjustment is typically necessary to compensate for the additional phase lag in the closed loop system due to the neglected dynamics of the probe. In this paper, we present a practical method for tuning of these parameters in AFM applications.

The stability and performance of the RC system are analyzed by the classical robust control approach [24], using the sensitivity, $S$, and complementary sensitivity, $T$, functions, which are defined in (3), where $L$ in (4) denotes the loop gain.

$$S = \frac{1}{1+L} \quad , \quad T = \frac{L}{1+L} \tag{3}$$

$$L = G\left(1 + \frac{q}{1 - qz^{-(n-nq)}} bz^{-(n-nq-nb)}\right)PI \tag{4}$$

The sensitivity function $S$ is required to be small for good tracking (i.e. disturbance rejection) at operating frequencies which are far below the resonance frequency of the stage, $\omega_{op} < \omega_{res}/10$, where the complementary function $T$ is required to be small for the frequencies around and above the resonance to achieve robust stability against unstructured modeling uncertainties (i.e. the neglected high frequency dynamics) and also to attenuate the sensor noise.

## III. Set-up

The sample to be scanned is placed on the piezo actuated *x-y-z* nano-stage (PI Inc., Germany, Model No. P-517.3CD), which has a travel range of 100×100×20 $\mu m^3$ and is equipped with integrated capacitive sensors for precise positioning (resolution: 0.1 nm). The stage is connected to a digital DAQ (data acquisition) card (PCI-DIO-96, National Instruments Inc.) via a parallel input/output port running a servo loop at 2 ms/cycle.

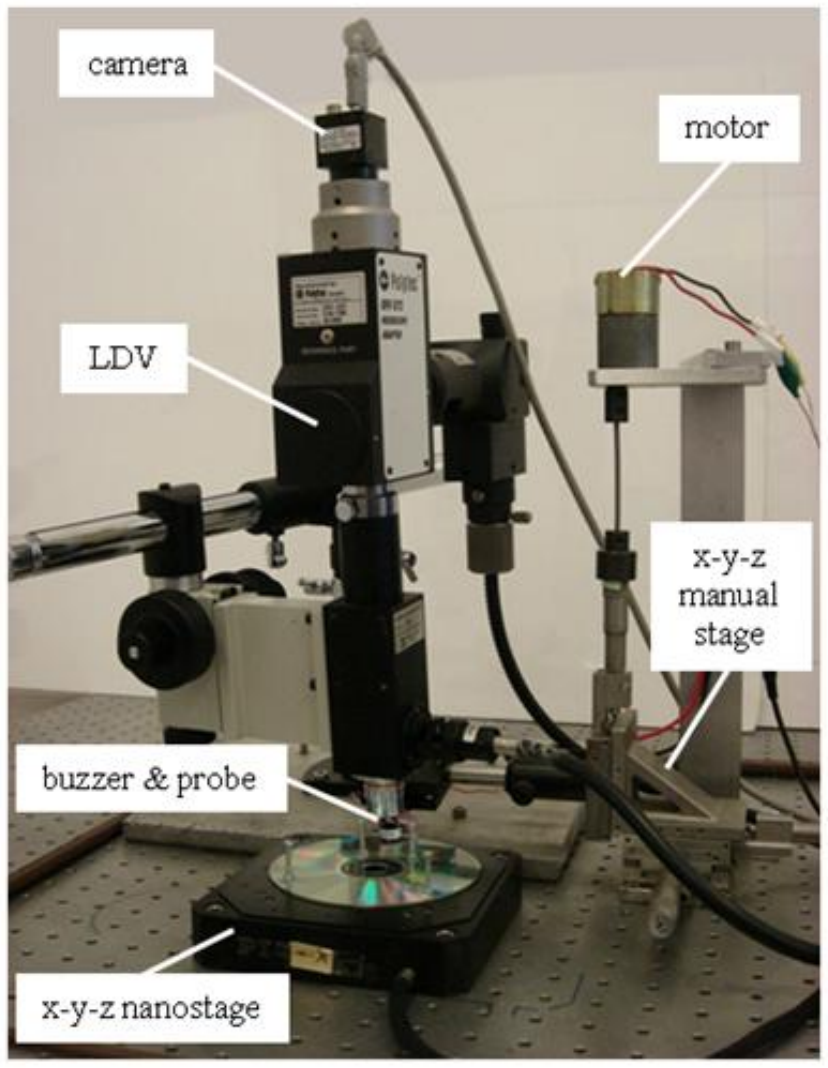

Fig. 3. The experimental setup for our tapping mode AFM.

The scanning probe used in the experiments is suitable for tapping mode AFM operation, and its resonance frequency is around 350 kHz (Olympus, OMCL-AC160TS). It is excited by a simple piezo-buzzer. The vibration of the probe at the tip is measured by a Laser Doppler Vibrometer (LDV) (Polytec GmbH, Germany). The output of the LDV is first sent to an RMS converter chip (Analog Devices, AD637JDZ) and then transferred to the computer through a separate DAQ unit (National Instruments USB 6251). The control signals for the raster scan motion on *x-y* plane and for imaging the surface along the *z* axis are transferred to the nano-stage through an amplifier. More details on the experimental setup (see Fig. 3) can be found in our earlier publication [25].

## IV. Identification of the Stage Dynamics

We used step and impulse signals as input to characterize the dynamic response of our nano-stage along the z-axis. A transfer function of the stage is constructed by the least squares system identification technique. The best fit is obtained for a transfer function having a polynomial of 3rd degree in the numerator and a polynomial of 7th degree in the denominator, as formulated below.

$$G(z)=\frac{b_3z^3+b_2z^2+b_1z+b_0}{z^7+a_6z^6+a_5z^5+a_4z^4+a_3z^3+a_2z^2+a_1z+a_0} \quad (5)$$

The coefficients of the polynomials are calculated using the method of least squares as

$$\underbrace{\zeta^{k+7}}_{\lambda}=\underbrace{\begin{bmatrix}-\zeta^{k+6} & \cdots & -\zeta^{k} & u^{k+3} & \cdots & u^{k}\end{bmatrix}}_{\Phi}\cdot\underbrace{\begin{bmatrix}a_6\\ \vdots\\ a_0\\ b_3\\ \vdots\\ b_0\end{bmatrix}}_{\theta}+e^k$$

for $k = 1\ldots N$ (6)

where, $u(z)$ and $\zeta(z)$ represent input and output functions respectively, $N$ is the number of sample points used for curve fitting, $\lambda$ is an $(N\text{-}7)\times 1$ vector, $\Phi$ is an $(N\text{-}7)\times 11$ matrix, and $\theta$ is a $11\times 1$ vector storing the constant coefficients. The coefficients minimizing the curve fitting error, $e$, can be calculated from the following relation

$$\theta=(\Phi^*\Phi)^{-1}\Phi^*\lambda. \quad (7)$$

Hence, the transfer function of the stage with the estimated coefficients is given in (8).

The actual step and the impulse responses of the stage obtained by the experiments are compared to the corresponding ones obtained from the transfer function model of the stage in Fig. 4. The frequency response of the stage obtained from the experimental impulse data via FFT (Fast Fourier Transform) is compared to that of the model in Fig. 5.

It is observed from Fig. 5 that the resonance frequency and the bandwidth of the stage are around 993Hz and 80 Hz, respectively. Using the magnitude and the phase diagrams of $G(z)$, one can calculate the gain margin of the stage as 3.55. Without causing instabilities, a gain value higher than this one can only be introduced by phase regulation as discussed further in the upcoming sections.

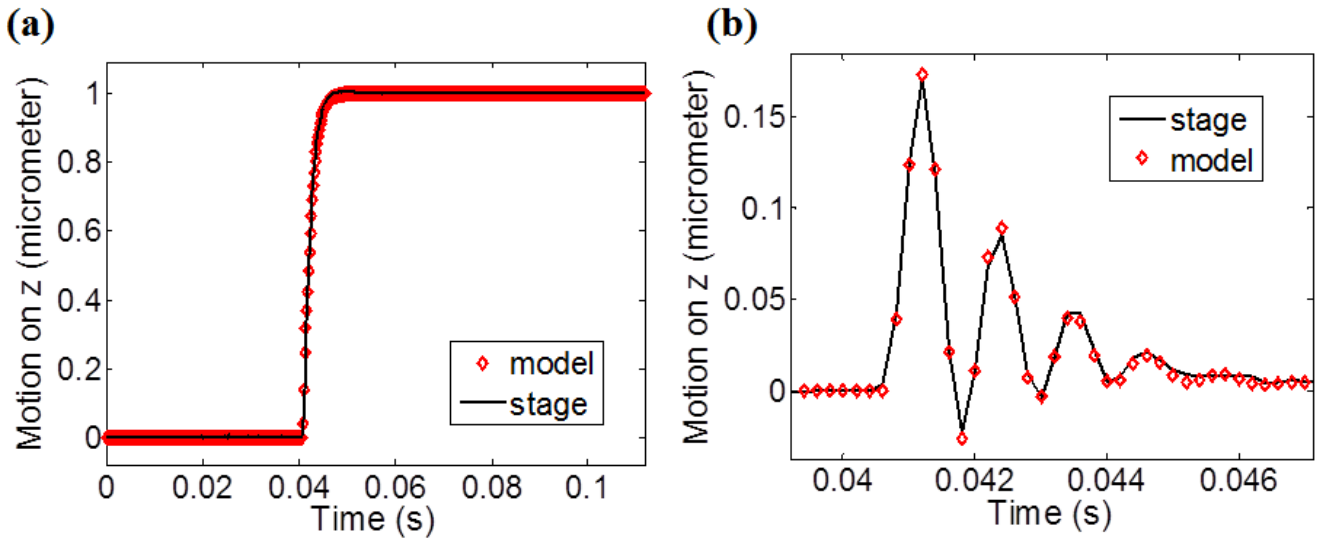

Fig. 4. The step (a) and the impulse (b) responses of the stage and the model.

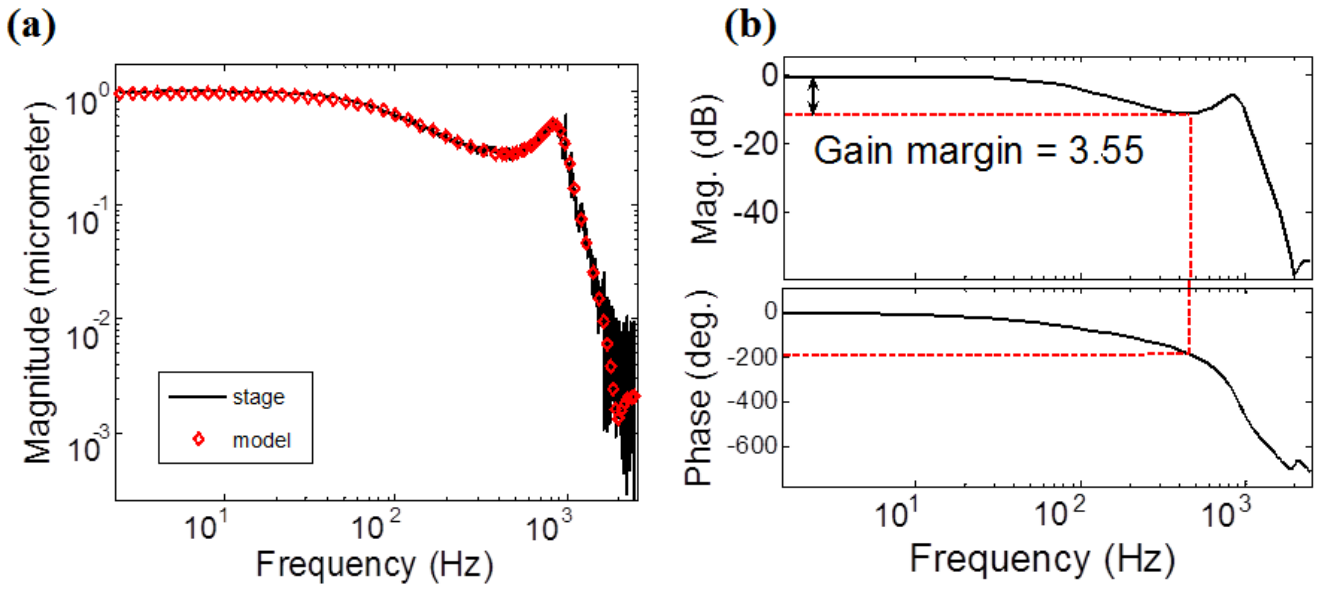

Fig. 5. The frequency response curves (a) and the Bode plots (b) of the stage and the model.

## V. Design and Analysis of the RC Filters

### A. *Design of the 'b' Filter*

We first tuned the PI controller shown in Fig. 2a for the best possible performance before adding the RC controller in series. Then, in order to make $R(\omega)<1$, the $b$ filter is chosen to be the inverse of $PI(z)G(z)/[1+PI(z)G(z)]$. As observed from (8), $G(z)$ has no non-minimum zeros, hence it can be inverted safely. Furthermore, $PI(z)G(z)/[1+PI(z)G(z)]$ is multiplied by a low-pass filter having a DC gain equal to one to satisfy the causality requirement. The poles of this low-pass filter are chosen to be distinct and close to the center of the unit circle. The resulting transfer function for $b$ filter is given in (9).

The effect of the $b$ filter on the dynamics of the system is shown in Fig. 6. Observe that the phase cannot be compensated well at frequencies higher than 100 Hz (see red colored solid curve). While a non-causal optimal inverse can be calculated for the $b$ filter to further improve the phase compensation as suggested in [26], this problem is handled by adding a small sample advance, $nb$, to the memory loop as already mentioned in Section II and will be further discussed in Section VI. The value of $nb = 4$ is estimated from the phase diagram given in Fig. 6a.

$$G(z)=\frac{0.0395z^3+0.0707z^2+0.0525z+0.0121}{z^7-1.3400z^6+1.1440z^5-0.4713z^4+0.0302z^3-0.1892z^2+0.1128z-0.1100} \quad (8)$$

$$b(z)=\frac{z^8-2.34z^7+2.484z^6-1.615z^5+0.554z^4-0.18z^3+0.294z^2-0.261z+0.098}{0.047z^8+0.045z^7-0.0077z^6-0.038z^5-0.012z^4+4.2\times10^{-8}z^3-5.1\times10^{-19}z^2} \quad (9)$$

### B. Design of the 'q' Filter

As mentioned earlier, the primary role of the $q$ filter is to reduce the negative effects of the noise in the system and the infinite gains introduced by the RC at the undesired harmonics of the input signal. For this reason, the cut-off frequency of the $q$ filter is chosen based on the mechanical bandwidth of the nano-stage (~ 80 Hz). Choosing a smaller value for the cut-off frequency would slow down the response of the RC while a higher value would cause the RC to replicate the undesired high frequency signals. Obviously, an ideal low-pass filter should have the gain value equal to one under the cut-off frequency, and zero above it. In order to converge to an ideal filter, we utilize a constant coefficient in the numerator of *q(z)* and a polynomial with a high degree in the denominator (see Eq. 10). Note that choosing a polynomial for the denominator higher than 4th degree does not significantly improve the performance of our controller. The phase introduced by *q(z)* is again compensated by adding a small sample advance, *nq*, to the memory loop (see Section VI). The value of $nq = 7$ is estimated from the phase diagram given in Fig. 6b.

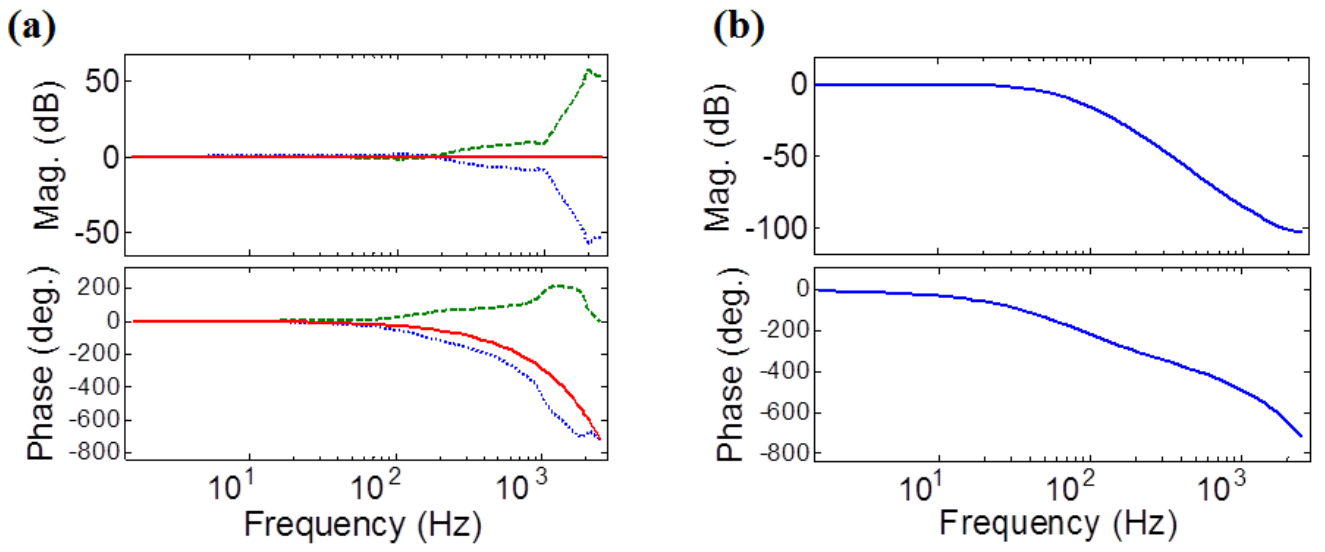

Fig. 6. (a) Bode plots of *b(z)* (green dashed), *PI(z)G(z)/[1+PI(z)G(z)]* (blue dotted), and *b(z)PI(z)G(z)/[1+PI(z)G(z)]* (red solid); (b) Bode plot of *q(z)*.

$$q(z) = \frac{8.853\times10^{-5}}{z^4 - 3.612z^3 + 4.892z^2 - 2.945z + 0.6649} \qquad (10)$$

### C. Analysis

It is desirable that a controller introduces high loop gains into the closed-loop system without causing instabilities. According to the robust design, the magnitude of the sensitivity function for the closed-loop system must be less than 1 at low frequencies, where the tracking performance is required, and that of the complementary sensitivity function must be less than 1 at high frequencies, where robust stability is required. Note that, it is not possible to satisfy the both requirements at the same time neither at low nor at high frequency. However, there is a transition region in between, where both requirements can be satisfied up to a certain degree.

For the analysis of tracking performance and robustness of the RC utilizing the filters designed in earlier sections, the loop gain, sensitivity and complementary sensitivity functions of the complete system (see Fig. 2a) are plotted and compared with that of the standalone PI controller in Figures 7, 8 and 9, respectively. The peaks in Fig. 7a are due to the high loop gains introduced by the RC at the fundamental frequency of 1 Hz and its harmonics (see blue dots). Fig. 8a shows that the closed loop bandwidth (i.e. control bandwidth) of the RC is two times larger than that of the stand-alone PI controller for the same frequencies. One can observe that the sensitivity of the RC is much lower than that of the stand-alone PI controller (see Fig. 8b), suggesting a better tracking performance. The complementary sensitivities of both controllers (the stand-alone PI and the RC) decrease rapidly after the resonance frequency to maintain their robustness against the unstructured modeling uncertainties (see Fig. 9a). Fig. 9b shows that the regeneration spectrum *R(ω)* is always less than one and the system is stable for all frequencies within the range of interest.

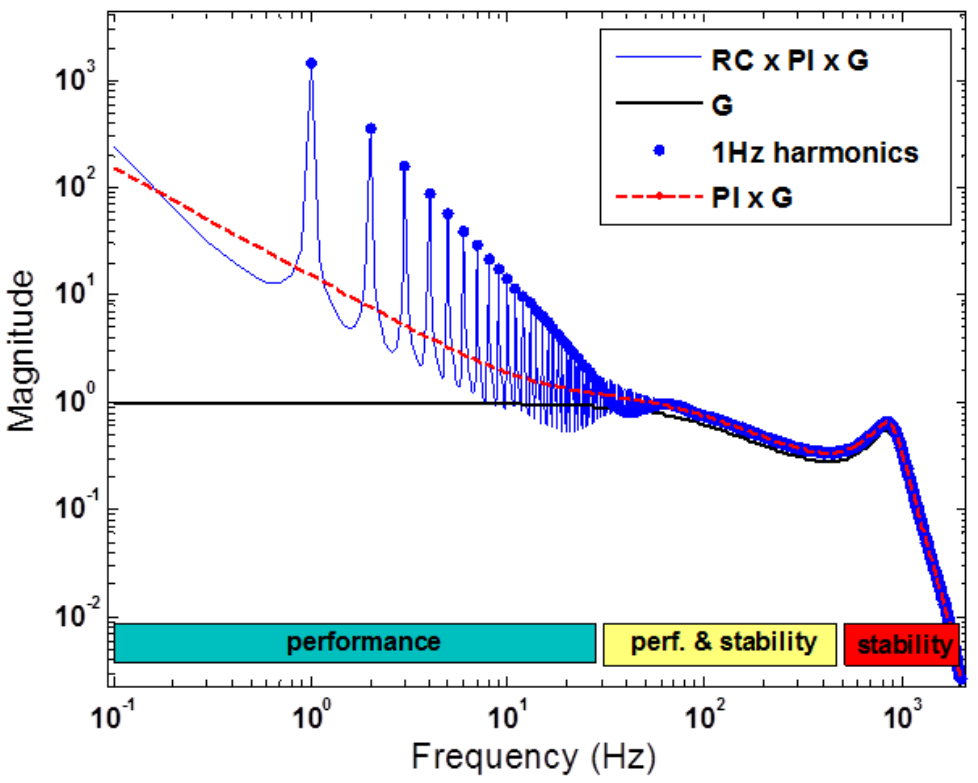

Fig. 7. Magnitude plots of the open-loop system including the stage, the stand-alone PI controller, and the RC.

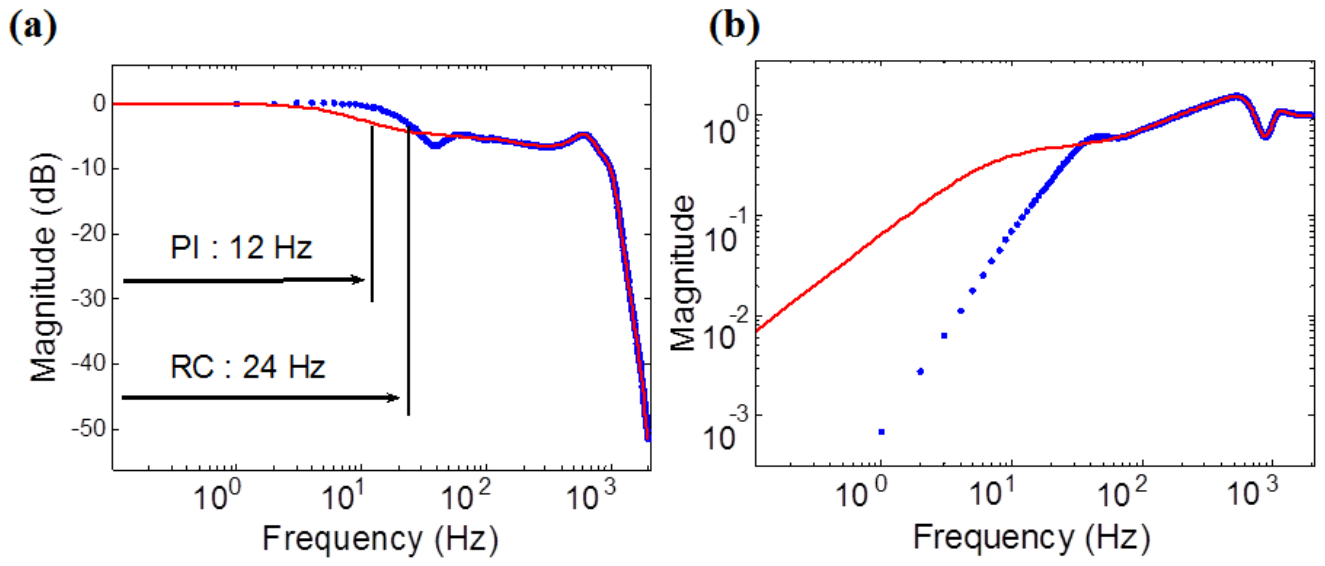

Fig. 8. (a) The magnitude plots of the closed-loop system under the stand-alone PI controller (red solid) and the RC (blue dotted). (b) The sensitivity functions of the closed-loop system under the stand-alone PI controller and the RC.

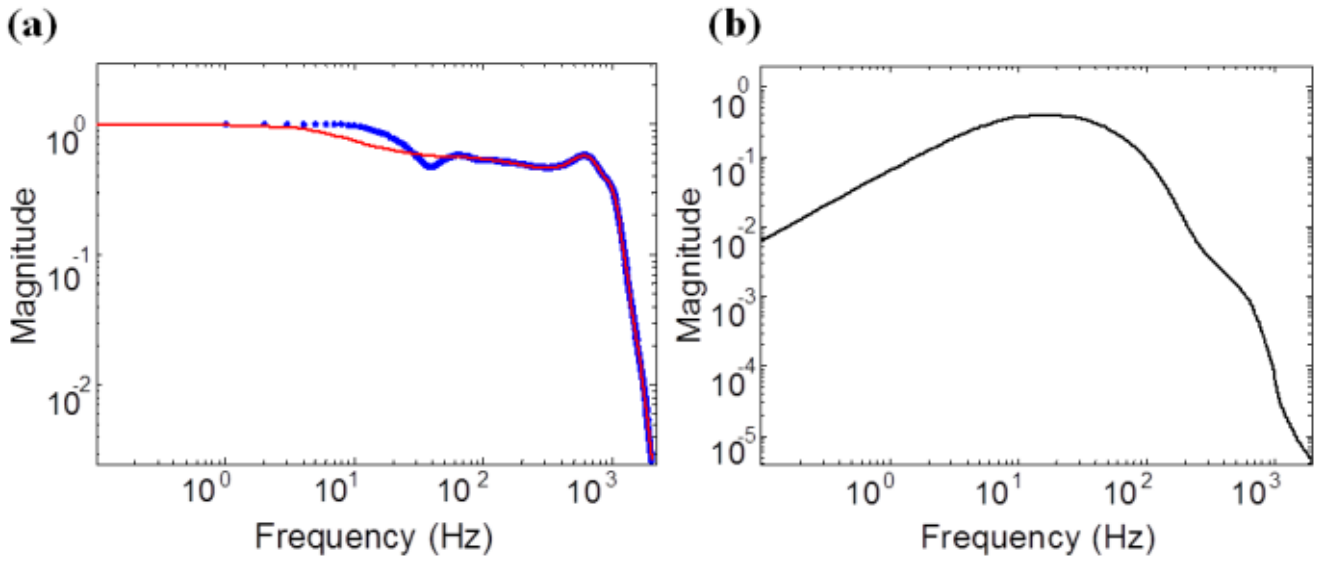

Fig. 9. (a) The complementary sensitivity functions of the closed-loop system under the stand-alone PI controller (red solid) and the RC (blue dotted.) (b) The reg neration spectrum of the closed-loop systems under the RC.

## VI. Experiments and Results

### A. Tracking Performance of the Stage

We first investigate the tracking performance of our stage along the *z*-axis by applying a square wave from a signal generator as the reference input. Hence, in this exercise, we do not consider the raster scan motion, the probe dynamics, and the force interactions between the probe and the sample at all. Our goal is to show that the RC improves the tracking performance of the stage significantly by reducing the rise time and compensating for the phase lag, which is not possible by using a conventional PI controller at high scan speeds.

The amplitude of the square wave used in the tracking experiments is 100 nm. The tracking experiments are performed at the wave frequencies of 1 Hz and 7 Hz. The proportional and integral gains of the stand-alone PI controller were set by trial and error for the wave frequency of 1 Hz. The sample advances for the RC are taken as $nb$ = 4 and $nq$ = 7 (see Section V). The results of the experiments show that the tracking performance of the stage under the RC is 49% and 66% better than that of the stand-alone PI controller for the wave frequencies of 1 Hz and 7 Hz, respectively (see Fig. 10). The tracking performance is quantified based on the difference between the reference input (i.e. square wave) and the stage output (i.e. the actual trajectory followed by the stage).

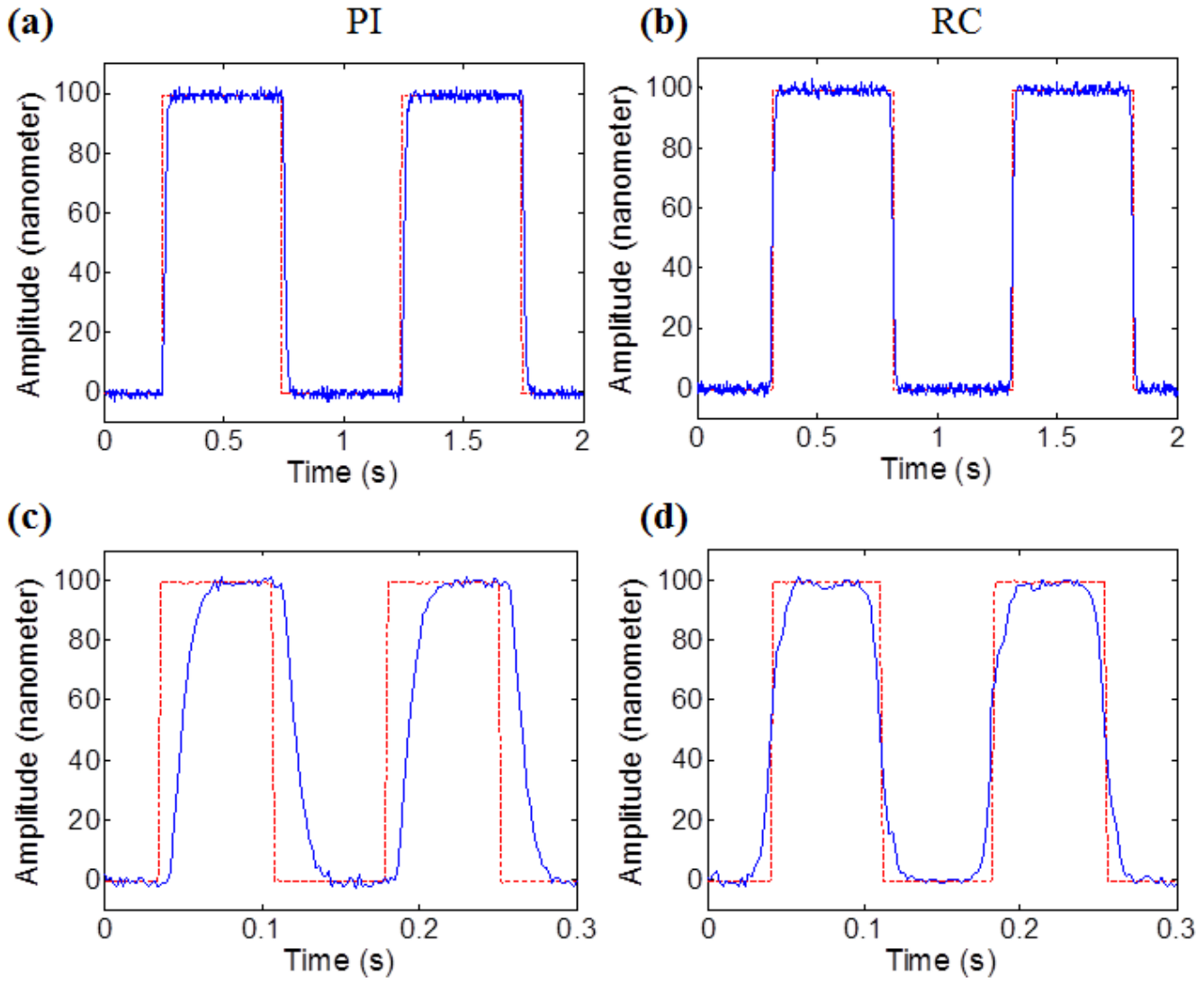

Fig. 10. The tracking performance of the stage under the stand-alone PI controller and the RC for the wave frequencies of 1 Hz (a and b, respectively) and 7 Hz (c and d, respectively).

The proposed RC assumes that the successive scan lines are similar and generates a control signal in the current scan line based on the error made in the previous scan line. To investigate the performance of the RC when the successive scan lines are not similar, two simple experiments are designed. In both experiments, a square wave of varying amplitude (frequency = 1 Hz) is tracked by the stage. For the sake discussion, we assume that each step in the wave represents one scan line. Hence, the RC memorizes the control signal generated in the previous step while scanning the current one. In the first experiment, the amplitude of the square wave is suddenly increased from 5 nm to 80 nm (see Fig. 11a and 11b). As shown in Fig. 11b, the RC has no difficulty in tracking the desired trajectory since the magnitude of the control signal kept in the memory loop of the RC to track the short steps is relatively small and dissolves in the larger control signal generated by the PI controller in series to the RC when a higher step is encountered. In the second experiment, the amplitude of the square wave being tracked is suddenly reduced from 80 nm to 5 nm. This time, we have observed some undesired peaks in the trajectory generated by the RC since the control signal carried on by the RC from the previous step is now much larger than the small value desired for the current step (see Fig. 11c and 11d). However, as shown in Fig. 11d, the undesired peaks quickly disappear after the second short step. Moreover, such a large difference (i.e. ~94%) between the profiles of two consecutive scan lines is unexpected in real AFM scans if sufficiently small advances are made along the $y$-axis. We also note that such undesired peaks do not appear at all when there is less than 50% reduction in the amplitude.

## B. AFM Scanning

The actual scanning experiments are conducted with a calibration grating having multiple steps (MicroMasch, TGZ02). The pitch and height of steps are 3.0 μm and 82.5 (±1.5) nanometers, respectively. The calibration grating is placed on the $x$-$y$-$z$ nano-stage. The tapping probe used in the experiments is initially brought close to the sample surface by means of a step motor (Fig. 3). A distance of a few hundred nanometers typically remains between the probe tip and the calibration grating after the initial adjustment. This initial distance is then compensated automatically by the $z$-axis controller of the stage at the very beginning of the scan process. The probe is excited by a simple piezo-buzzer to vibrate with an amplitude of $A_{free}$ = 52 nm in the free air, where the set amplitude is chosen as $A_{set} = 0.7\ A_{free}$.

In our initial testing of the RC for scanning the grating,

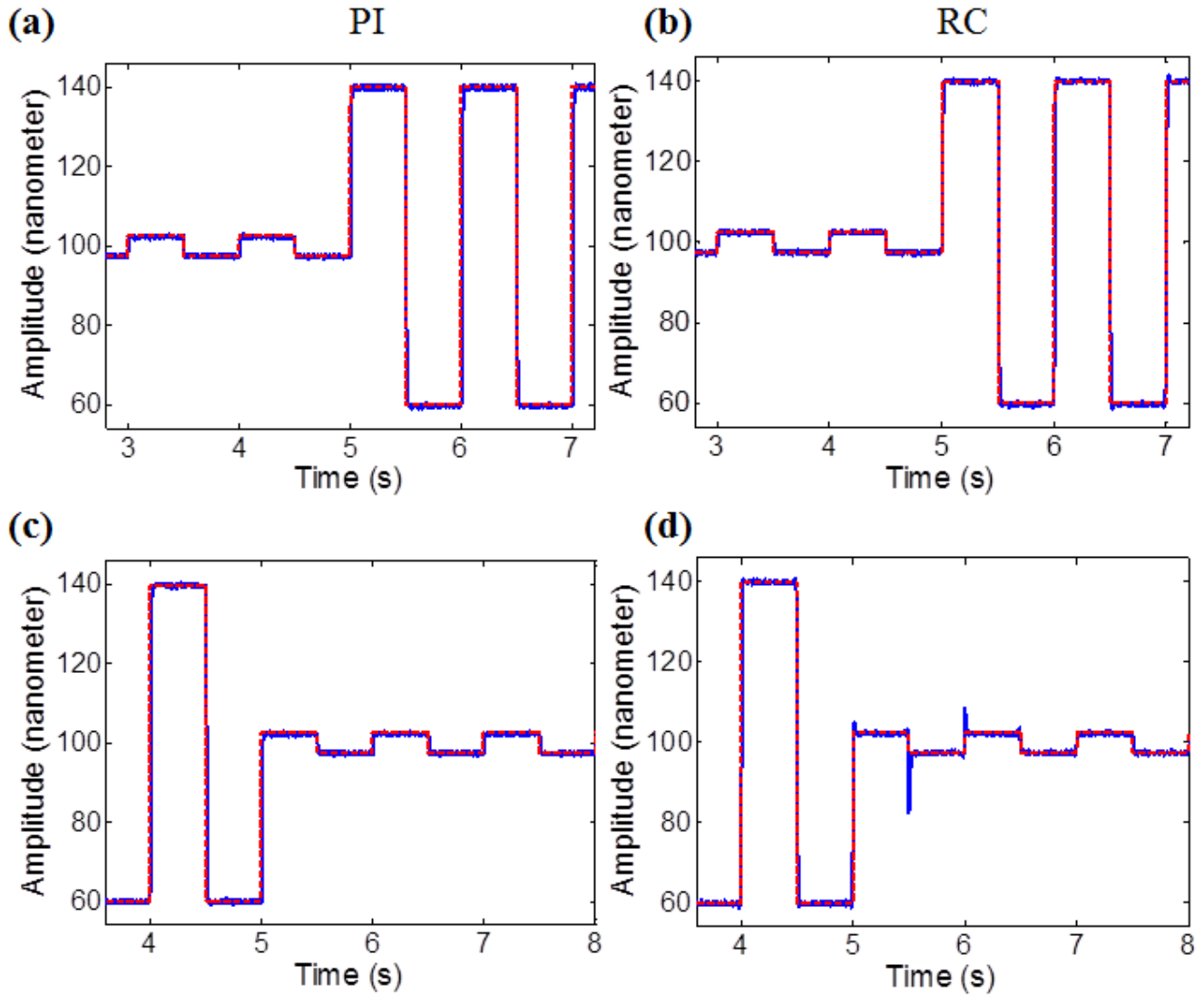

Fig. 11. The tracking performance of the stage under the stand-alone PI controller and the RC for the wave frequencies of 1 Hz when a large fluctuation occurs in the amplitude (For the sake discussion, we assume that each step represents one scan line).

sudden peaks in the error signal ($A_{set}$ – $A_{act}$) are observed. They indicate an additional phase lag in the overall response of the system due to the neglected dynamics of the probe and the nonlinear force interactions between the probe tip and the scanned surface. To compensate for this additional lag, the small sample advances (*nb* and *nq*), initially estimated from the phase diagrams in Section II, are slightly altered to achieve the desired performance. We developed an iterative approach to estimate the optimal values of *nb* and *nq* resulting in the best performance. For this purpose, a new error measure, named as cumulative scan error (CSE), is defined to quantify the scan performance of the RC for the different combinations of *nb* and *nq* within a range. The CSE is calculated by using the error signal recorded during the scan, *e(t)*, the scan speed, *v*, and the sampling time, $T_s$, and then normalized by the step dimensions as

$$CSE = \frac{\sum |e(t)| T_s v}{h w} . \qquad (11)$$

The range of *nb* and *nq* values tested for the best performance is varied from 1 to 10 and 1 to 20 respectively, based on their initial estimates (*nb* = 4 and *nq* = 7). In our iterative approach, first, *nb* is set to the initial value of 4 and the scan experiments are performed for each integer value of *nq*. Then, the value of *nq* returning the minimum scan error is chosen to search for the optimal value of *nb* using the same approach. After several iterations, the optimal values of *nb* and *nq* are estimated as 5 and 9, respectively (see Fig. 12). As shown in Fig. 12, the scan error is more sensitive to variations in *nb* than *nq*.

Following the setting of *nb* and *nq*, more comprehensive scanning experiments at different scan speeds are performed and the performance of the RC is compared to that of the stand-alone PI controller using the CSE. The results presented in this paper are for the lowest and the highest scan speeds of 3 µm/s and 21 µm/s (see Figures 13 and 14, respectively). Since the pitch of the calibration steps used in the scan experiments is 3 µm, these scan speeds correspond to the wave frequencies of 1 Hz and 7 Hz used in the experiments performed with the stage in the previous section. The results show that the CSE calculated for the RC is 30 % less than that of the stand-alone PI controller at the scan speed of 3 µm/s. Although the scan profiles obtained by the RC and the stand-alone PI controller appear to match the desired profile very well, the phase delay in PI controller (see the peaks appearing in the error signal in Fig. 13b) results larger CSE than that of the RC. This time delay further increases and causes even larger CSE for the stand-alone PI controller as the scan speed is increased to 21 µm/s. Compared to the stand-alone PI controller, the relative improvement achieved by the RC in scan profile was 66 % at the scan speed of 21 µm/s (Fig. 14).

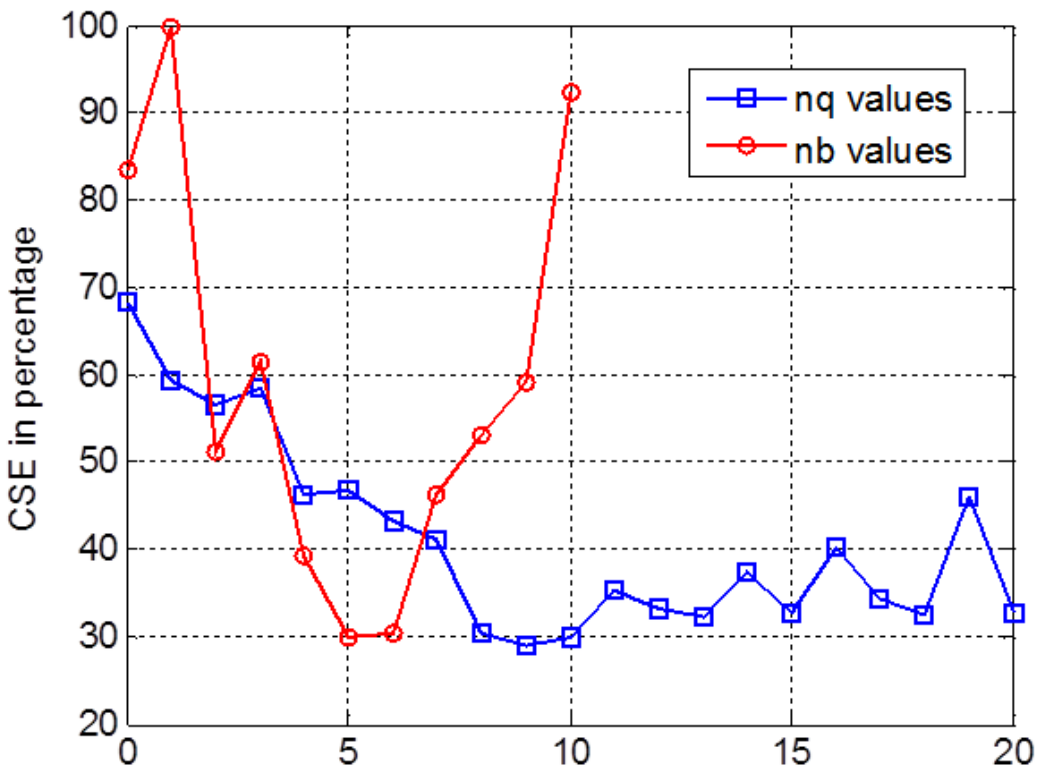

Fig. 12. The variation of *nb* and *nq* as a function of CSE (normalized).

In addition to the tracking performance, the tapping forces observed in the RC are compared to that of the stand-alone PI controller. It is reported in [27] that the average tapping forces can be estimated as $\langle F \rangle = (A_{free}^2 - A_{set}^2)^{1/2} k/Q_{eff}$, where *k* is the stiffness of the cantilever probe and $Q_{eff}$ is its effective quality factor. To estimate the instantaneous tapping forces, the set amplitude is replaced by the instantaneous one in the above formula. In our set-up, *k* = 42 N/m and $Q_{eff}$ = 436 for the probe and the set amplitude is 70 % of the free air amplitude (i.e. 52 nm), which results in a nominal tapping force of 4 nN (see the red colored horizontal lines in Figures 13c and 14c). Ideally, the controller should maintain the interaction forces between the probe tip and the sample surface at this nominal value for perfect imaging. As the tapping forces deviate from the nominal value, the image quality is reduced. The results show that the deviation of the tapping forces from the nominal value is small when the RC is used for scanning (compared to the stand-alone PI controller, a reduction of 40 % and 60 % is achieved for the scan speeds of 3 μm/s and 21 μm/s, respectively). Moreover, the average tapping forces above the nominal value are reduced by 5.5% and 58% for the scan speeds of 3 μm/s and 21 μm/s, respectively. In order to reduce the tracking errors and the tapping forces, the RC takes an early control action in the current scan line based on the control signal generated for the previous scan line.

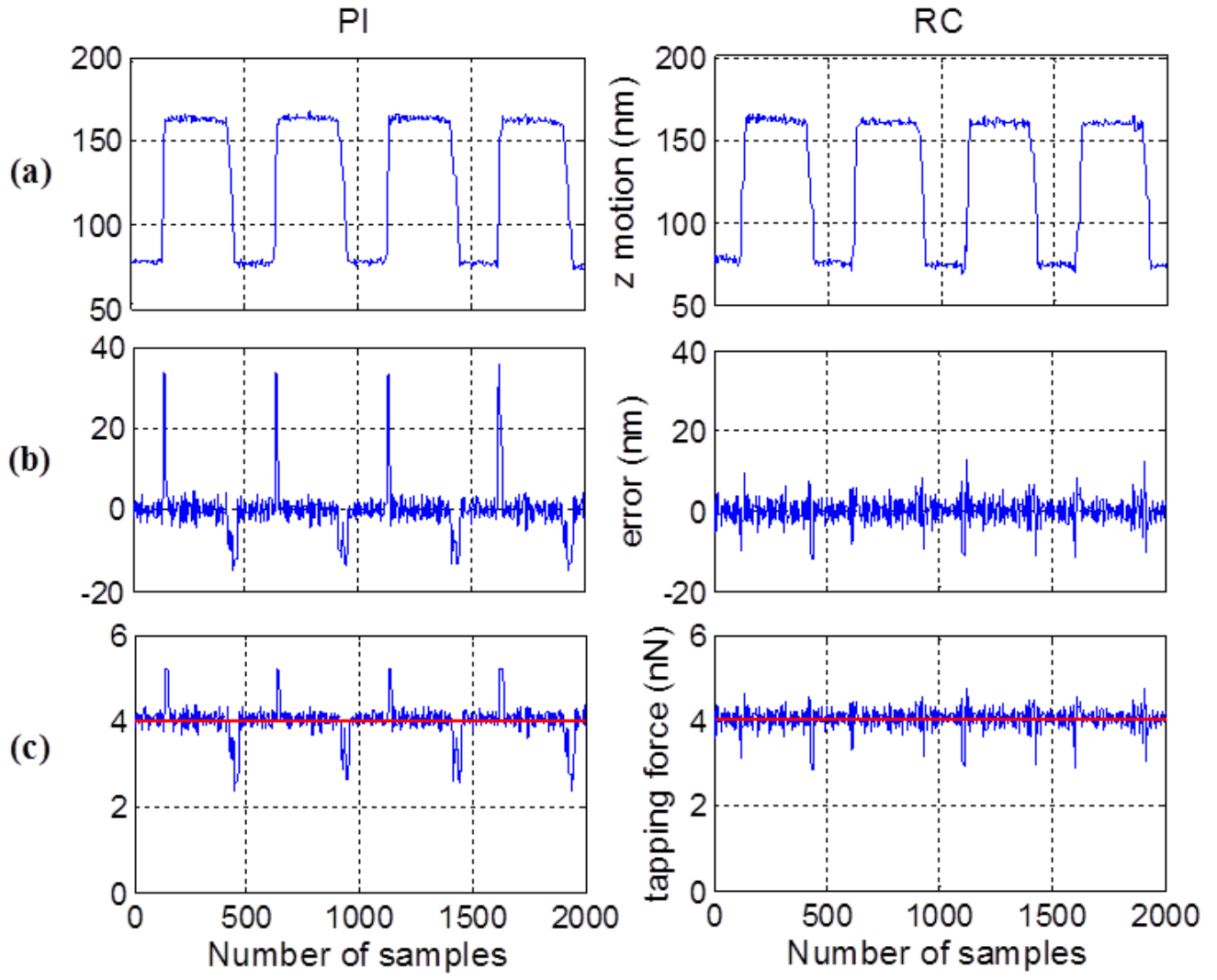

Fig. 13. The scan results for the stand-alone PI controller (left column) and the RC (right column) at the scan speed of 3 µm/s: (a) scan profiles, (b) error signal, (c) tapping forces.

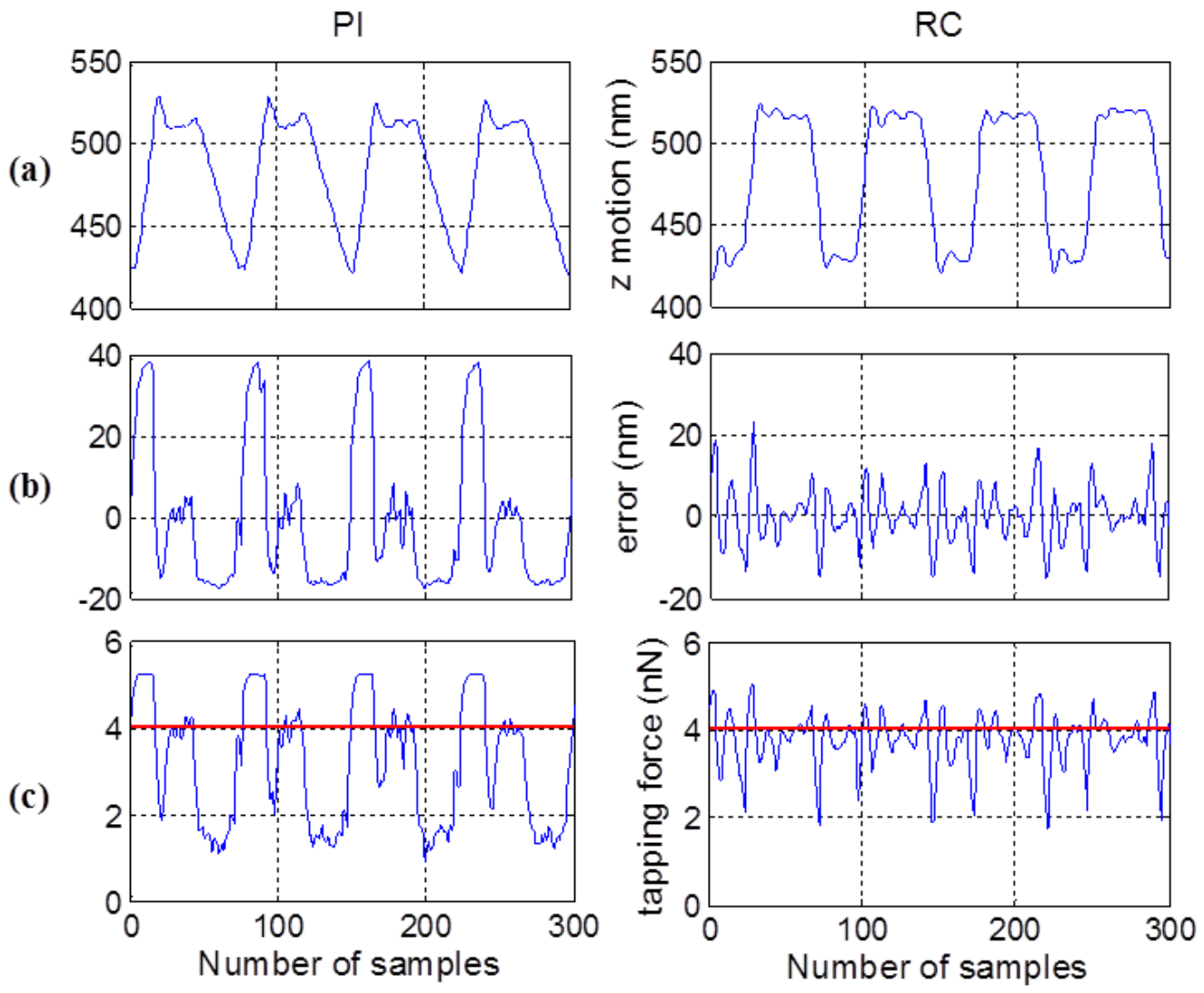

Fig. 14. The scan results for the stand-alone PI controller (left column) and the RC (right column) at the scan speed of 21 µm/s: (a) scan profiles, (b) error signal, (c) tapping forces.

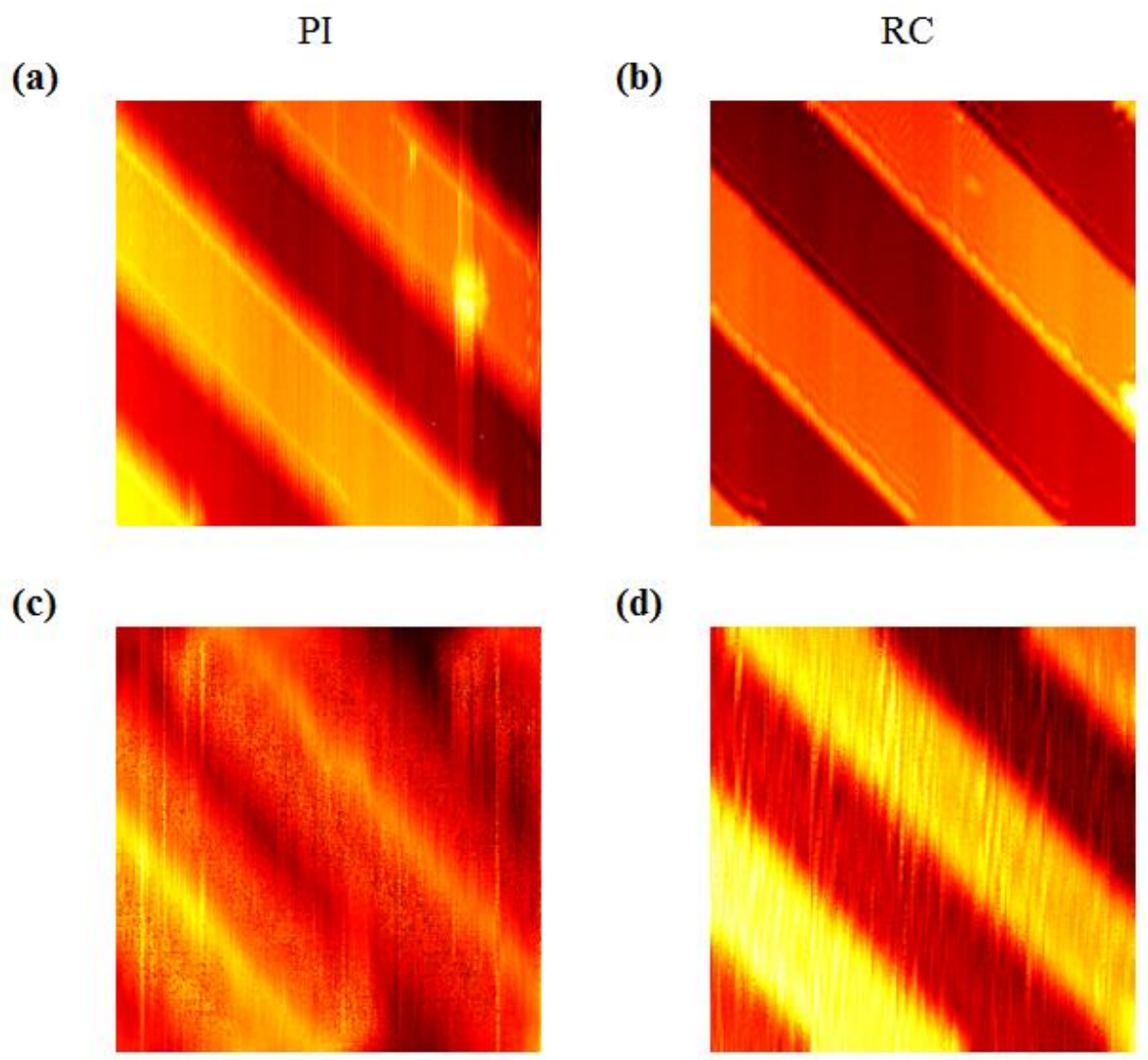

Fig. 15. The scan results for the stand-alone PI controller (left column) and the RC (right column) at the scan speeds of 3 μm/s (a and b) and 21 μm/s (c and d) when the calibration grating is rotated 45 degrees. The scanned area is 6 μm × 6 μm.

To investigate the performance of the RC when the phase of the repetitive signal changes, we rotated the calibration grating by 45 degrees about the z-axis and scanned the resulting diagonal lines. Obviously, a slight phase difference occurs between the consecutive scan lines in this configuration as the stage advances along the *y*-axis. As shown in Fig. 15, the tracking errors due to this phase difference are compensated well by the RC controller, but not by the stand-alone PI controller (Fig. 15).

## VII. Discussion and Conclusion

We implemented an RC to achieve better performance than the conventional PI controller in scanning nano-scale surfaces at high scan speeds without sacrificing from the image quality. Moreover, we showed that the magnitude of the tapping forces between the scanning probe and the surface was significantly decreased by the RC. For example, when the scan speed was increased by 7-fold, the scan error and the average tapping forces were reduced by 66% and 58%, respectively. The reduction in the magnitude of tapping forces is important since high values can easily damage the specimen (especially, the biological ones) as well as the scanning probe.

The scan performance achieved by the RC is better than the stand-alone PI controller since the successive scan lines are similar and the proposed RC takes into account the knowledge of the previous scan line while scanning the current one. The similarity of the successive scan lines is certainly true for a sample having a smooth surface profile but also holds for a rougher surface since the topography being measured is smoothened by the probe tip due to its finite radius and cone angle [12]. In the stand-alone PI controller, when the probe encounters the rising edge of a step, the vibration amplitude of the probe reduces to zero rapidly and sticking occurs, which appears as a sudden peak in the error signal (see Figures 13b and 14b). The sticking is almost unavoidable since the control signal calculated in the current servo loop based on the error signal is only effective in the next servo loop. On the other hand, the RC can take a control action in the current scan line based on the error estimated for the previous scan line. As a result, we have less sticking, smaller error, and smaller tapping forces compared to the stand-alone PI controller. When the probe encounters the falling edge of the same step, it suddenly detaches from the sample surface and the magnitude of the error cannot exceed $A_{set}-A_{free}$, which limits the speed of the response and hence the tracking performance of the stand-alone PI controller. It takes longer time for the probe oscillations to reach the desired amplitude $A_{set}$ again. Hence, the resulting scan profile during the saturation period is erroneous. On the other hand, the error signal entering to the PI controller shown in Fig 2a is augmented by the RC and hence the error saturation is eliminated and a faster response is achieved (Fig. 13a).

While the performance of the RC is better than that of the stand-alone PI controller in many aspects, one must know the period of the repetition for its implementation. In our case, this period is the time spent for the back and forth motions of the stage along a single scan line. Since the user sets the period of this raster scan motion, the above requirement can be satisfied easily. Hence, the proposed RC only requires that the successive scan lines to be repetitive, but their profiles do not have to be periodic. We should also mention that the implementation of the RC requires small sample advances (*nb* and *nq*) in the memory loop to compensate for the phase delays introduced by the filters *b* and *q* and the closed-loop system itself. Their selection is critical for achieving better performance, especially at high scan speeds. However, we showed that the iterative approach proposed in Section VI could successfully determine the optimal values of *nb* and *nq* once the good initial estimates were provided. Those initial values can easily be obtained from the phase diagrams of the associated filters.

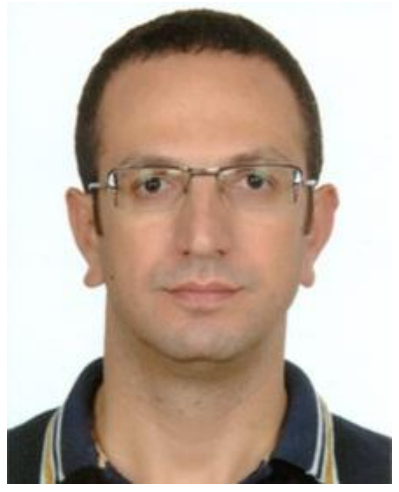

**Serkan Necipoglu** is currently a graduate student in the Mechanical Engineering Program of Istanbul Technical University, Istanbul, Turkey. Serkan has BS degree in Mechanical Engineering from Yildiz Technical University, Istanbul, Turkey in 2002 and MS degree in Control Systems from Imperial College London, UK in 2004. His research interests include control systems, system dynamics, mechatronics, and nanotechnology. Contact him at necipoglu@itu.edu.tr.

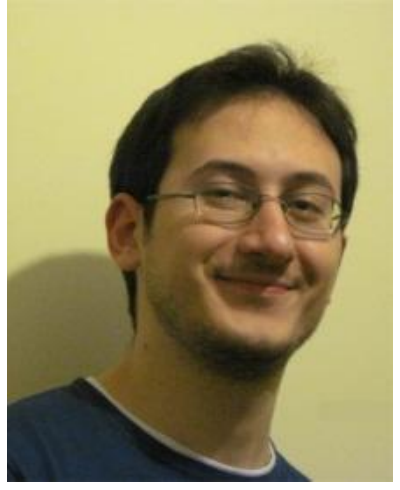

**Selman A. Cebeci** is a graduate student in the Mechanical Engineering Department of Koc University, Istanbul, Turkey. Selman has BS degree in Mechanical Engineering from Sabanci University, Istanbul, Turkey in 2008. His research interests include control systems, mechatronics, and nanotechnology. Contact him at secebeci@ku.edu.tr.

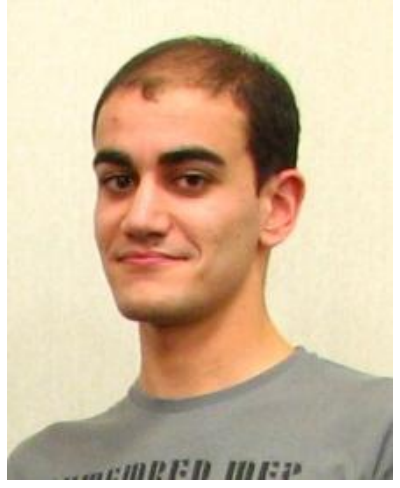

**Yunus E. Has** is a graduate student in the Mechanical Engineering Department of Koc University, Istanbul, Turkey. Yunus has BS degree in Mechanical Engineering from Baskent University, Ankara, Turkey in 2008. His research interests include control systems, mechatronics, and nanotechnology. Contact him at yhas@ku.edu.tr.

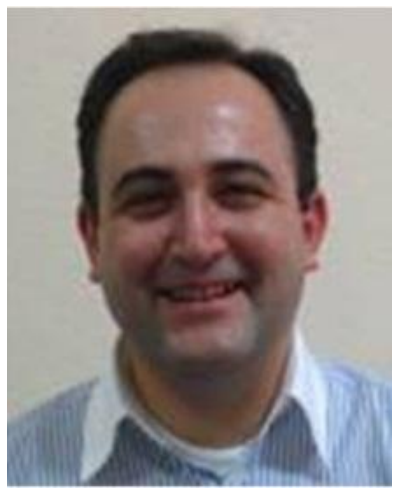

**Levent Guvenc** is a faculty member in the Mechanical Engineering Department of Istanbul Technical University, Istanbul, Turkey. He is the director of the Automation Lab and the EU funded Automotive Controls and Mechatronics Research Center. Guvenc received the B.S. degree in Mechanical Engineering from Bogazici University in Istanbul, in 1985, the M.S. degree in Mechanical Engineering from Clemson University in 1988, and the Ph.D. degree in Mechanical Engineering from the Ohio State University in 1992. His current research interests are automotive control and mechatronics, helicopter stability and control, mechatronics and applied robust control. Contact him at guvencl @itu.edu.tr.

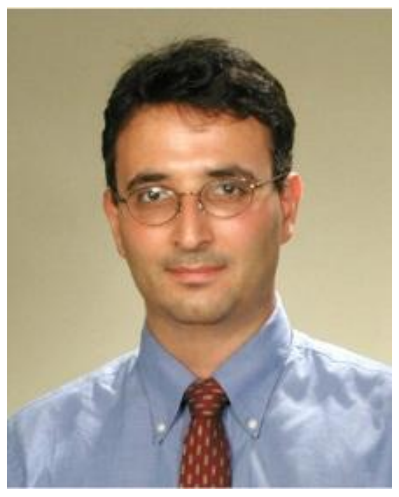

**Cagatay Basdogan** is a faculty member in the Mechanical Engineering and Computational Sciences and Engineering programs of Koc University, Istanbul, Turkey. He is also the director of the Robotics and Mechatronics Laboratory at Koc University. Before joining to Koc University, he worked at NASA-JPL/Caltech, MIT, and Northwestern University Research Park. His research interests include haptic interfaces, robotics, mechatronics, biomechanics, medical simulation, computer graphics, and multi-modal virtual environments. Basdogan has a PhD in Mechanical Engineering from Southern Methodist University in 1994. He is currently the associate editor of IEEE Transactions on Haptics and Computer Animation and Virtual Worlds journals. Contact him at cbasdogan@ku.edu.tr.